\documentclass[reprint]{revtex4-2}
\usepackage{import, preamble}
\usepackage{float}

\begin{document}
\pagestyle{plain}

\title{Interplay Between Network Heterogeneity and Activation Thresholds in Global Cascades}

\author{Aidan Meador-Woodruff}
\email{ameadorw@ur.rochester.edu}
\affiliation{Department of Physics and Astronomy,
University of Rochester, Rochester, New York 14627, USA}

\author{Gourab Ghoshal}
\email{gghoshal@pas.rochester.edu}
\affiliation{Department of Physics and Astronomy,
University of Rochester, Rochester, New York 14627, USA}
\affiliation{Department of Computer Science,
University of Rochester, Rochester, New York 14627, USA}

\begin{abstract}
The onset of global cascades in complex networks depends
on both the distribution of activation thresholds and the
underlying network topology.
Here we investigate how these two sources of heterogeneity
interact within the Watts threshold model, using beta-distributed
activation thresholds and beta-binomial degree distributions
to systematically vary their respective properties.
Analyzing the onset of cascades from infinitesimal seeds,
we identify two distinct boundaries governing the
large-connectivity behavior of vulnerable branching.
The first separates regimes in which increasing connectivity
enhances or suppresses propagation, depending on the
abundance of individuals with low activation thresholds.
The second distinguishes two mechanisms of suppression:
one controlled by the scarcity of low-threshold individuals
and the other by the diminishing abundance of low-degree
nodes as connectivity increases.
At the boundary between these mechanisms, their competing
contributions produce a logarithmic correction to the
asymptotic scaling.
Our results demonstrate how the interplay between
network structure and individual activation thresholds
determines the conditions for extensive propagation.
\end{abstract}

\maketitle

\section{Introduction}

The propagation of small perturbations into system-wide
events is a characteristic feature of many complex systems.
Examples range from cascading failures in electrical grids
and interdependent infrastructure to financial contagion,
where the failure of individual components can trigger
widespread disruptions
\citep{MotterLai2002,Yang2017,Brummitt2012,GaiKapadia2010}.
Similar phenomena arise in biological and social systems,
including neuronal avalanches
\citep{BeggsPlenz2003,ShewPlenz2013},
the emergence of superspreading events in epidemiology
\citep{LloydSmith2005},
and the propagation of behaviors, opinions, and information
through social networks
\citep{Centola2010,CentolaMacy2007,Monsted2017,DelVicario2016}.
In the context of epidemics, recent studies have also
demonstrated how the organization of population density
and mobility flows can influence the susceptibility of
urban systems to disease outbreaks
\citep{Hazarie2021,Aguilar2022}.
While the mechanisms underlying these processes differ,
a common question concerns the conditions under which
local interactions and initially small perturbations can
produce large-scale collective responses.

A particularly simple framework for addressing this
question is the threshold model of collective behavior
introduced by Granovetter \citep{Granovetter1978}
and subsequently extended to random networks by
Watts \citep{Watts2002}.
In this model, individuals or components are represented
by nodes on a network and become active when a sufficient
fraction of their neighbors have activated.
The resulting dynamics depend on both the underlying
network topology and the distribution of activation
thresholds.
Watts showed that the propagation of infinitesimal
perturbations is governed by the connectivity of vulnerable
nodes, those that can be activated by a single active
neighbor.
This leads to a characteristic feature of threshold
dynamics: increasing network connectivity can initially
facilitate global cascades but eventually suppress them,
as highly connected nodes become increasingly resistant
to activation by individual neighbors.
The resulting cascade window provides a simple explanation
for why intermediate levels of connectivity can be
particularly susceptible to extensive propagation.

The original framework has been extended to account for
finite seed fractions, heterogeneous network structures,
and more general response functions
\citep{Gleeson2008,GleesonPorter2018}.
In particular, Gleeson and Cahalane developed an analytical
treatment of cascade dynamics for finite initial seed
fractions \citep{GleesonCahalane2007}.
Related work has demonstrated how degree heterogeneity
affects connectivity, robustness, and percolation in
complex networks \citep{AielloChungLu2000,Albert2000},
including the structure of bicomponents
\citep{NewmanGhoshal2008}, responses to heterogeneous
failures \citep{KarrerGhoshal2008}, and percolation
transitions \citep{MimarGhoshal2022}.
Threshold heterogeneity introduces a distinct source
of variability, since nodes with different activation
thresholds respond differently to the same fraction
of active neighbors \citep{Watts2002,GleesonCahalane2007}.
These studies establish the importance of both network
structure and individual response, but leave open how
the two interact in determining the onset of global
cascades.

We therefore examine how these two sources of
heterogeneity interact when both are varied systematically.
The distinction between them is not merely one of
network structure versus individual response.
A node's degree determines both the number of neighbors
through which it can propagate activation and the fraction
of those neighbors that must become active before it
responds.
Consequently, the contribution of a particular threshold
distribution to cascade propagation depends on how
vulnerability is distributed across nodes of different
degrees.
For example, a population containing many individuals
with low activation thresholds need not support extensive
cascades if the network does not connect those individuals
sufficiently to sustain propagation.
Conversely, increasing connectivity can provide additional
pathways for activation while simultaneously reducing
the vulnerability of individual nodes.
Understanding how these competing effects depend on
the shapes of the degree and threshold distributions
is therefore central to characterizing the onset and
suppression of global cascades.

This question is also relevant to social contagion,
where the adoption of behaviors or opinions often
requires reinforcement from multiple contacts
\citep{CentolaMacy2007,Centola2010}.
Online social networks provide a particularly interesting
setting because both the patterns of interaction and
the information to which individuals are exposed can
change over time.
Recommendation algorithms, for instance, can influence
the visibility of information and the structure of
social interactions, potentially affecting both the
opportunities for exposure and the reinforcement required
for an individual to engage
\citep{piccardi_reranking_2025}.
Although such processes involve additional complications,
including correlations between individual behavior and
network structure, they provide motivation for understanding
the separate and combined effects of degree and threshold
heterogeneity.

Here we investigate this question within the global
cascade model by introducing a continuously tunable
family of degree and threshold distributions.
For activation thresholds, we consider the beta
distribution, whose shape parameters allow us to vary
the population of low-threshold individuals while
prescribing the mean threshold.
In particular, its behavior near zero is controlled
by a single exponent, providing a direct way to examine
how the abundance of vulnerable nodes affects cascade
propagation.
Previous analytical studies have demonstrated how
different network evolution rules generate heterogeneous
degree distributions \citep{MooreGhoshalNewman2006}
and how prescribed degree distributions can be obtained
through suitable local growth mechanisms
\citep{GhoshalNewman2007}.
For the network topology, we use the beta-binomial
distribution, a natural extension of the binomial
distribution that incorporates heterogeneity in
connection probabilities \citep{Bakbergenuly2017}.
Unlike the commonly studied Poisson and power-law
distributions, it allows us to examine a broad range
of degree heterogeneity within a finite-support family,
without invoking diverging moments.
Furthermore, its representation as a mixture of
binomial distributions provides a convenient analytical
route for studying the cascade condition.
Together, these choices allow us to vary the two
sources of heterogeneity systematically and identify
the mechanisms controlling their interaction.
We consider uncorrelated, locally tree-like random
networks with independently assigned thresholds
and examine the cascade condition in the
infinitesimal-seed limit.

Our analysis identifies two distinct boundaries
governing the large-connectivity behavior of
vulnerable branching.
The first occurs at the threshold shape parameter
$\alpha_{\rm th}=1$, separating regimes in which
increasing connectivity enhances or suppresses
the propagation of infinitesimal perturbations.
Interestingly, a related boundary has recently
been identified in the original Granovetter
threshold model with beta-distributed thresholds
\citep{Fontanari2026}.
The second boundary,
$\alpha_{\rm th}=\alpha_d+2$,
arises from the competition between the degree
and threshold distributions.
Below this boundary, the asymptotic behavior
is controlled by the lower tail of the threshold
distribution; above it, the dominant contribution
comes from the diminishing abundance of low-degree
nodes.
At the boundary itself, the competition between
these contributions produces a logarithmic correction.
These results characterize how the two sources
of heterogeneity jointly determine the
large-connectivity behavior of the cascade condition.

We begin by revisiting the cascade condition for
representative combinations of degree and threshold
distributions, before introducing the beta and
beta-binomial families and examining their effects
separately.
We then consider the doubly tunable model and
derive the asymptotic regimes governing the
interaction between threshold and degree heterogeneity.
We conclude with a discussion of the physical
interpretation of these results, their relationship
to previous work on cascade dynamics, and the
limitations and possible implications of the framework.

\section{Model and Analysis}
\label{sec:fourcases}

We consider an uncorrelated, locally tree-like random network
with degree distribution $p_k$, where $p_k$ is the probability
that a node has degree $k$, and mean degree
$z=\langle k\rangle=\sum_k k p_k$.
Each node is independently assigned an activation threshold
$\phi$ drawn from a cumulative distribution $F(\phi)$,
independent of its degree.
A node becomes active when the fraction of its active
neighbors reaches or exceeds its threshold. Once activated,
nodes remain active.

For an infinitesimal fraction of initially active nodes,
the onset of global cascades is governed by the propagation
of activation through vulnerable nodes, defined as those
that can be activated by a single active neighbor.
A node of degree $k$ is therefore vulnerable with
probability $F(1/k)$.

Following a randomly chosen edge reaches a node of degree
$k$ with probability $kp_k/z$. If this node is vulnerable,
it can propagate activation through its $k-1$ remaining
connections. The corresponding branching parameter is
\begin{equation}
    \kappa
    =
    \frac{1}{z}
    \sum_{k=2}^{\infty}
    k(k-1)p_k
    F\left(\frac{1}{k}\right),
    \label{eq:kappa_intro}
\end{equation}
which is the mean reproduction number of the vulnerable
branching process.

The condition $\kappa>1$ marks the percolation threshold
of the vulnerable subnetwork~\citep{Watts2002,GleesonCahalane2007}.
Above this threshold, a giant vulnerable component exists,
and an infinitesimal seed can trigger a global cascade
with nonzero probability. This condition establishes
the possibility of extensive propagation, but does not
determine the probability that a particular seed triggers
a cascade or its final size.

The vulnerable-cluster condition should be distinguished
from the existence of an ordinary giant connected component,
which requires
$\langle k(k-1)\rangle/z>1$
but does not account for node vulnerability.
A network may therefore possess a giant connected component
without supporting a giant vulnerable component.

Equation~\eqref{eq:kappa_intro} provides the starting point
for our analysis. Both the degree distribution $p_k$ and
the threshold distribution $F$ enter the same branching
condition. Their interaction determines whether the
vulnerable population can sustain extensive propagation.

\subsection{Motivating examples}
\label{sec:motivating_examples}

\begin{figure*}[t!]
\centering
\includegraphics[width=\linewidth]{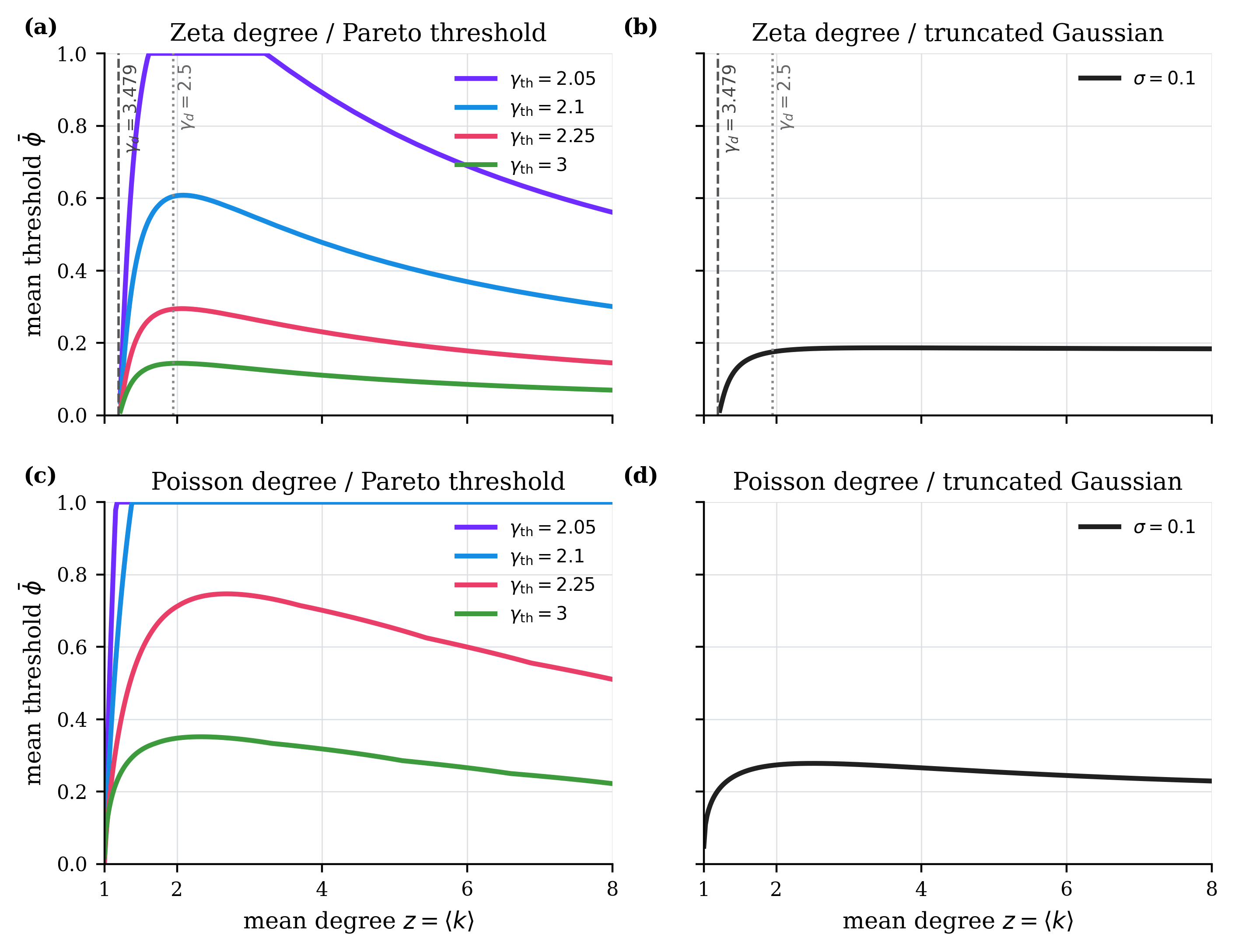}
\caption{Contours of the vulnerable-cluster
condition $\kappa=1$ in the $(z,\bar\phi)$ plane
for four combinations of degree and threshold
distributions: (a) zeta degrees and Pareto
thresholds, (b) zeta degrees and truncated Gaussian
thresholds, (c) Poisson degrees and Pareto
thresholds, and (d) Poisson degrees and truncated
Gaussian thresholds. Pareto threshold exponents
are $\gamma_{\rm th}=2.05,2.10,2.25,3.00$;
the truncated Gaussian has underlying scale
parameter $\sigma=0.1$. The vertical lines
in the upper panels mark $\gamma_d=2.5$
and the structural percolation threshold
$\gamma_d^*\approx3.479$. The contours
separate regions satisfying the
infinitesimal-seed vulnerable-cluster
condition $\kappa>1$ from those that do not.
The vertical coordinate is the actual mean
threshold after truncation.}
\label{fig:four-cases-kappa}
\end{figure*}

To illustrate how degree and threshold heterogeneity
jointly affect the onset of global cascades, we first
consider four combinations of degree and threshold
distributions. For the network topology, we compare
Poisson degree distributions, characteristic of
Erd\H{o}s--R\'enyi networks, with power-law degree
distributions of the zeta form,
\begin{equation}
    p_k^{\mathrm{ER}}
    =
    e^{-z}\frac{z^k}{k!},
    \qquad
    p_k^{\mathrm{PL}}
    =
    \frac{k^{-\gamma_d}}{\zeta(\gamma_d)},
    \quad k\geq 1.
\end{equation}
For the zeta distribution, the mean degree is
$z=\zeta(\gamma_d-1)/\zeta(\gamma_d)$, which is
finite for $\gamma_d>2$. Unlike the Poisson case,
the mean degree is therefore fixed by the exponent
$\gamma_d$.

For the threshold distribution, we first consider
a Pareto distribution with cumulative distribution
function
\begin{equation}
    F_{\mathrm{P}}(x)
    =
    \begin{cases}
        0, & x<x_{\min},\\[1mm]
        1-\left(\dfrac{x_{\min}}{x}\right)^{\gamma_{\rm th}-1},
        & x\geq x_{\min},
    \end{cases}
\end{equation}
where
\begin{equation}
    x_{\min}
    =
    \bar\phi\,
    \frac{\gamma_{\rm th}-2}{\gamma_{\rm th}-1},
    \qquad \gamma_{\rm th}>2.
\end{equation}
This choice fixes the mean threshold at $\bar\phi$.
As $\gamma_{\rm th}\to2^+$, the lower cutoff
approaches zero and an increasing fraction of
the threshold probability becomes concentrated
near zero. The distribution can also assign
thresholds greater than one, corresponding to
nodes that cannot activate under the fractional
threshold rule.

For comparison, we consider a Gaussian threshold
distribution truncated and normalized on $[0,1]$.
Writing $\varphi$ and $\Phi$ for the standard normal
density and cumulative distribution function,
respectively, the truncated CDF is
\begin{equation}
    F_{\mathrm{G}}(x)
    =
    \begin{cases}
        0, & x<0,\\[1mm]
        \displaystyle
        \frac{
        \Phi\left(\frac{x-\mu}{\sigma}\right)
        -
        \Phi\left(-\frac{\mu}{\sigma}\right)}
        {
        \Phi\left(\frac{1-\mu}{\sigma}\right)
        -
        \Phi\left(-\frac{\mu}{\sigma}\right)},
        & 0\leq x\leq1,\\[3mm]
        1, & x>1.
    \end{cases}
\end{equation}
Here $\mu$ and $\sigma$ are the location and
scale parameters of the underlying Gaussian.
Because truncation changes the mean, the actual
mean threshold is
\begin{equation}
\begin{aligned}
    \bar\phi
    &=
    \mu+\sigma
    \frac{
    \varphi(-\mu/\sigma)
    -
    \varphi((1-\mu)/\sigma)}
    {
    \Phi((1-\mu)/\sigma)
    -
    \Phi(-\mu/\sigma)}.
\end{aligned}
\end{equation}
We use this mean when constructing the cascade
contours, rather than identifying $\bar\phi$
with the underlying Gaussian location $\mu$.

For each combination, we evaluate the
vulnerable-cluster condition
$\kappa=1$ using Eq.~\eqref{eq:kappa_intro}.
Figure~\ref{fig:four-cases-kappa} shows the
resulting contours in the $(z,\bar\phi)$ plane.

The threshold distribution strongly influences
the region supporting a supercritical vulnerable
cluster. For Pareto thresholds, decreasing
$\gamma_{\rm th}$ toward $2$ expands this region,
reflecting the increasing concentration of
threshold probability near zero. However, for
any fixed $\gamma_{\rm th}>2$, the Pareto
distribution has a positive lower cutoff.
Consequently, sufficiently high-degree nodes
cannot be activated by a single neighbor.
The expansion of the cascade region must
therefore be distinguished from the singular
limit $\gamma_{\rm th}\to2^+$.

The degree distribution imposes a separate
constraint. For the zeta distribution with
minimum degree one, an ordinary giant
connected component exists only for
$\gamma_d<\gamma_d^*\approx3.479$.
Beyond this point, extensive cascades
triggered by infinitesimal seeds are
impossible, regardless of the threshold
distribution. Within the connected regime,
the zeta topology generally supports a
smaller region satisfying $\kappa>1$
than the corresponding Poisson topology
for the parameters shown.

The Gaussian examples illustrate a different
aspect of threshold heterogeneity. For a
truncated Gaussian with nonzero density
$f_{\mathrm{G}}(0)$ at zero, the probability
of single-neighbor vulnerability scales as
$F_{\mathrm{G}}(1/k)\sim f_{\mathrm{G}}(0)/k$
at large degree. Consequently, on Poisson
networks the cascade parameter approaches
$\kappa_\infty=f_{\mathrm{G}}(0)$ as
$z\to\infty$. Whether vulnerable branching
remains supercritical therefore depends on
the density of thresholds near zero,
rather than simply on the width of the
Gaussian distribution.

Together, these examples show that cascade
onset depends not simply on whether the degree
or threshold distribution is heterogeneous,
but on how their respective distributions
determine the abundance and connectivity
of vulnerable nodes. This motivates the
continuously tunable beta and beta-binomial
families considered in the following sections.

\subsection{Heterogeneous Activation Thresholds}
\label{sec:betath}

\begin{figure*}[t!]
\centering
\includegraphics[width=\linewidth]{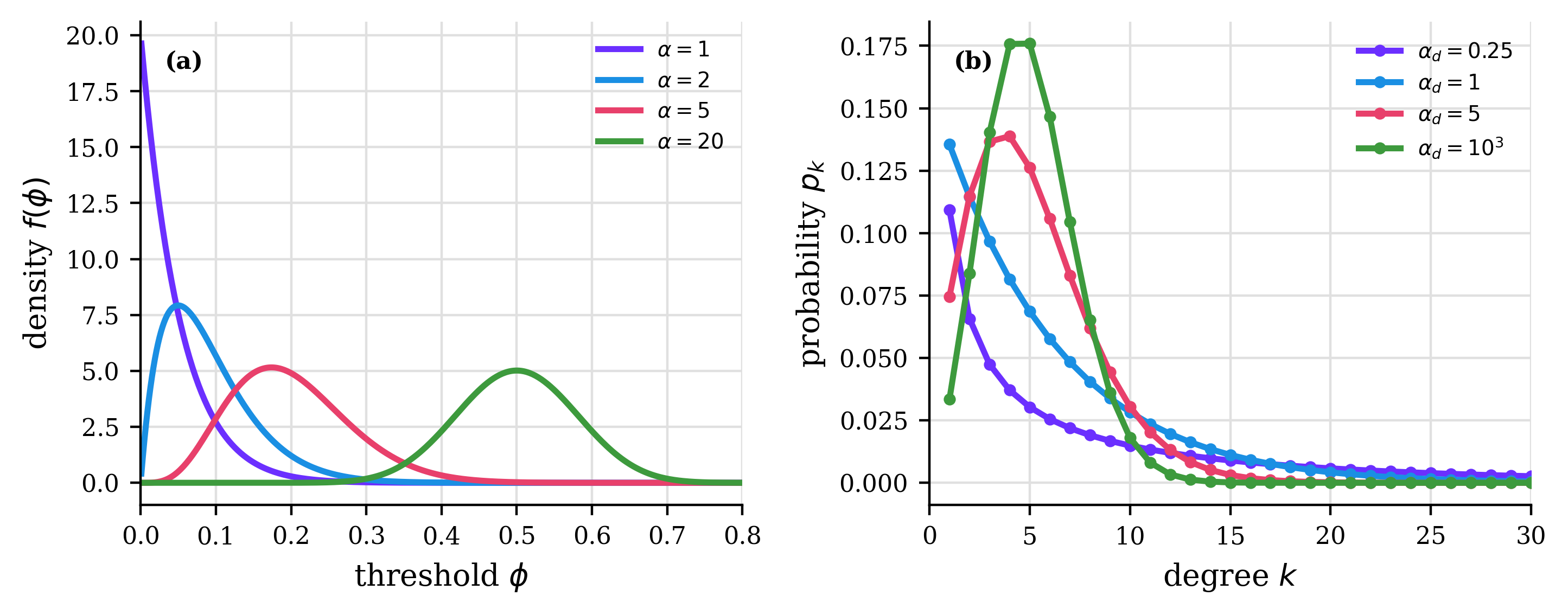}
\caption{Examples of the beta and beta-binomial
distributions at varying levels of heterogeneity.
(a) shows beta threshold distributions
with parameters $\mathrm{Beta}(1,20)$,
$\mathrm{Beta}(2,20)$, $\mathrm{Beta}(5,20)$,
and the symmetric $\mathrm{Beta}(20,20)$.
These are the unscaled distributions; the cascade
calculations use rescaled beta distributions with
fixed mean $\bar\phi$.
(b) shows beta-binomial degree
distributions with mean degree $\langle k\rangle=5$
for $\alpha_d=0.25,1,5,1000$ with
$\beta_d=5,20,95,10^5$, illustrating the progression
from strongly skewed to increasingly concentrated
degree distributions.}
\label{fig:beta-dist}
\end{figure*}

We begin by considering a threshold distribution that can be
varied continuously between strongly heterogeneous and
narrowly concentrated forms. We use the beta distribution,
whose probability density is
\begin{equation}
    f(x;\alpha_{\rm th},\beta_{\rm th})
    =
    \frac{x^{\alpha_{\rm th}-1}(1-x)^{\beta_{\rm th}-1}}
    {B(\alpha_{\rm th},\beta_{\rm th})},
\end{equation}
where $B(\alpha,\beta)=\Gamma(\alpha)\Gamma(\beta)/
\Gamma(\alpha+\beta)$.
The parameter $\alpha_{\rm th}$ controls the behavior near
zero, with $f(x)\sim x^{\alpha_{\rm th}-1}$ as $x\to0$.
For $\alpha_{\rm th}<1$, the density diverges at zero,
producing an enhanced population of nodes with very low
activation thresholds. At $\alpha_{\rm th}=1$, the density
approaches a finite, nonzero value, while for
$\alpha_{\rm th}>1$, it vanishes at zero. When
$\alpha_{\rm th}=\beta_{\rm th}$, the distribution is
symmetric about $1/2$. Examples of these distributions
are shown in the left panel of Fig.~\ref{fig:beta-dist}.

To vary threshold heterogeneity while holding the mean
threshold fixed, we rescale the beta distribution.
If $X\sim\mathrm{Beta}(\alpha_{\rm th},\beta_{\rm th})$,
with mean
$m=\alpha_{\rm th}/(\alpha_{\rm th}+\beta_{\rm th})$,
we define $\Phi=(\bar\phi/m)X$, so that
$\langle\Phi\rangle=\bar\phi$.
The corresponding cumulative distribution function is
\begin{equation}
    F(x)=
    \begin{cases}
    I_{mx/\bar\phi}(\alpha_{\rm th},\beta_{\rm th}),
        & 0\leq x<\bar\phi/m,\\
    1, & x\geq\bar\phi/m,
    \end{cases}
\end{equation}
where $I_u(\alpha,\beta)$ is the regularized incomplete
beta function,
\begin{equation}
    I_u(\alpha,\beta)
    \equiv
    \frac{1}{B(\alpha,\beta)}
    \int_0^u t^{\alpha-1}(1-t)^{\beta-1}\,dt.
\end{equation}
This rescaling fixes the mean while allowing the shape
of the threshold distribution to vary. It can, however,
produce thresholds greater than one, corresponding to
nodes that cannot activate under the fractional-threshold
rule. We retain this convention throughout.

We first examine the effect of threshold heterogeneity
on Erd\H{o}s--R\'enyi networks, whose degree distribution
is Poisson with mean $z=\langle k\rangle$.
Substituting the Poisson degree distribution into
Eq.~\eqref{eq:kappa_intro} gives
\begin{equation}
    \kappa(z)
    =
    \sum_{k=2}^{\infty}
    \frac{k(k-1)p_k}{z}
    F\left(\frac{1}{k}\right),
\end{equation}
where $p_k=e^{-z}z^k/k!$.
The contour $\kappa=1$ marks the onset condition
for a supercritical vulnerable cluster in the
infinitesimal-seed limit.

In the large-connectivity limit $z\to\infty$,
the Poisson distribution becomes concentrated around
$k\approx z$.
Using the near-zero expansion of the incomplete beta
function, Eq.~(A4) of Appendix A, we have
\begin{equation}
    F(1/k)\sim C_{\rm th}k^{-\alpha_{\rm th}},
    \label{eq:er-threshold-tail}
\end{equation}
where
$C_{\rm th}
=m^{\alpha_{\rm th}}/
[\alpha_{\rm th}B(\alpha_{\rm th},\beta_{\rm th})
\bar\phi^{\alpha_{\rm th}}]$. Consequently, the cascade parameter scales as
\begin{equation}
    \kappa(z)
    \sim
    C_{\rm th}z^{1-\alpha_{\rm th}}.
    \label{eq:er-asym}
\end{equation}

\begin{figure}[t]
\centering
\includegraphics[width=\linewidth]{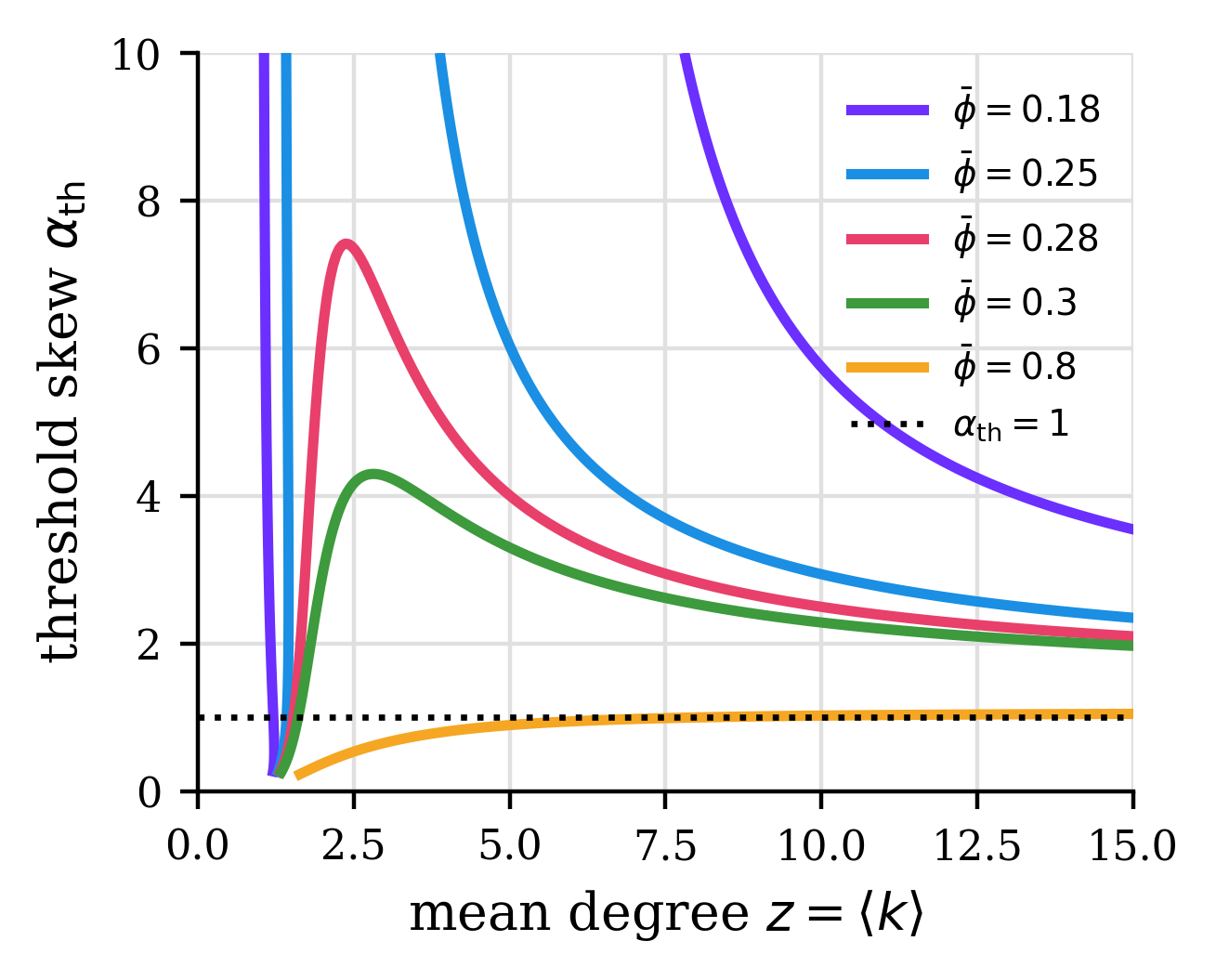}
\caption{Contours of the vulnerable-cluster condition
$\kappa=1$ in the $(z,\alpha_{\rm th})$ plane for
Erd\H{o}s--R\'enyi networks at different mean
thresholds $\bar\phi$.
The dotted horizontal line marks
$\alpha_{\rm th}=1$, separating regimes of increasing
and decreasing vulnerable branching in the
large-connectivity limit.
The colored contours delimit parameter regions
satisfying $\kappa>1$ and illustrate the dependence
of finite-connectivity cascade windows on the
mean activation threshold.
The condition $\kappa>1$ establishes the possibility
of extensive propagation from infinitesimal seeds,
but does not specify the probability or final size
of a cascade.}
\label{fig:beta-er-kappa}
\end{figure}

The exponent $1-\alpha_{\rm th}$ determines whether
increasing connectivity enhances or suppresses
vulnerable branching. For $\alpha_{\rm th}<1$,
$\kappa$ grows with $z$, and sufficiently connected
networks satisfy the global-cascade onset condition
$\kappa>1$. Although individual nodes become less
likely to activate through a single neighbor as
their degrees increase, the number of potential
secondary activations grows faster than this
vulnerability declines. For $\alpha_{\rm th}>1$,
the opposite occurs: vulnerability decreases faster
than connectivity increases, and the vulnerable
branching process eventually becomes subcritical.
These contrasting behaviors are illustrated by
the contours in Fig.~\ref{fig:beta-er-kappa}.

At $\alpha_{\rm th}=1$, the two effects balance,
and the cascade parameter approaches a finite
constant. This limiting behavior is not restricted
to Erd\H{o}s--R\'enyi networks. For
$\alpha_{\rm th}=1$, the threshold CDF satisfies
\begin{equation}
    F(1/k)
    =
    \frac{C}{k}+O(k^{-2}),
    \qquad
    C=\frac{\beta_{\rm th}m}{\bar\phi}.
\end{equation}
Substituting into the general cascade condition,
we obtain
\begin{equation}
\begin{aligned}
    \kappa(z)
    &=
    \frac{C}{z}
    \sum_{k=2}^{\infty}(k-1)p_k
    +O(z^{-1})\\
    &=
    C+O(z^{-1}),
\end{aligned}
\end{equation}
where the second equality follows from
$\sum_{k=2}^{\infty}(k-1)p_k=z-1+p_0$.
Thus, for any family of degree distributions
whose mean $z$ grows without bound, with a
fixed threshold distribution independent of
degree, the cascade parameter approaches
\begin{equation}
    \kappa_\infty
    =
    \lim_{z\to\infty}\kappa(z)
    =
    \frac{\beta_{\rm th}}
    {(1+\beta_{\rm th})\bar\phi}.
    \label{eq:kappa-infinity}
\end{equation}
The result is independent of the degree
distribution, without requiring Poisson
statistics or assumptions about higher moments.

Setting $\kappa_\infty=1$ defines the critical
mean threshold
\begin{equation}
    \bar\phi_{\rm crit}
    =
    \frac{\beta_{\rm th}}{1+\beta_{\rm th}}.
    \label{eq:phi-crit}
\end{equation}
For $\bar\phi<\bar\phi_{\rm crit}$, the
vulnerable branching process remains supercritical
at arbitrarily high connectivity. For
$\bar\phi>\bar\phi_{\rm crit}$, it becomes
subcritical in the large-connectivity limit.
At equality, the limiting cascade parameter
is unity, and the leading asymptotic result
alone does not determine the finite-connectivity
behavior.

The finite-connectivity behavior can be more
complicated than these asymptotic limits suggest.
For $\alpha_{\rm th}>1$, the cascade parameter
vanishes as $z\to\infty$, while it is also
subcritical at sufficiently small connectivity.
Consequently, when $\kappa$ exceeds unity at
intermediate $z$, a bounded cascade window
emerges, as in Watts' original model.
Increasing the mean threshold generally reduces
the population of vulnerable nodes and can
shrink or eliminate this window.
The disappearance of a finite-connectivity
window, however, is not determined solely by
the large-$z$ exponent and need not coincide
with $\alpha_{\rm th}=1$.

The boundary at $\alpha_{\rm th}=1$ is therefore
an asymptotic result rather than a universal
condition for the disappearance of a finite-$z$
cascade window. It identifies the balance
between increasing connectivity and decreasing
single-neighbor vulnerability.
A related threshold exponent appears in
Granovetter's model with beta-distributed
thresholds~\cite{Fontanari2026}.
As we show in Sec.~\ref{sec:doubly_tunable},
the same boundary governs the high-connectivity
behavior of the beta-binomial family, although
the scaling away from $\alpha_{\rm th}=1$
depends on the degree distribution.

\subsection{Heterogeneous Network Topology}
\label{sec:bbtopo}

We now turn to a tunable degree distribution. While
power-law degree distributions provide a useful model
of heterogeneous topology, varying their exponent
also changes their moments and imposes restrictions
on the accessible mean degree. To examine the effects
of degree heterogeneity more systematically, we use
the beta-binomial distribution, obtained by mixing
binomial distributions with a beta-distributed
success probability. This provides a continuous
family of degree distributions ranging from strongly
skewed to narrowly concentrated forms.

The beta-binomial probability mass function is
\begin{equation}
    P(k;n,\alpha_d,\beta_d)
    =
    \binom{n}{k}
    \frac{B(k+\alpha_d,n-k+\beta_d)}
    {B(\alpha_d,\beta_d)},
    \label{eq:bb-pmf}
\end{equation}
where $n$ is the maximum possible degree and
$\alpha_d,\beta_d>0$ are shape parameters.
We distinguish $n$ from the number of network
vertices $N$.

The mean and variance are
\begin{align}
z=\langle k\rangle
&=n\frac{\alpha_d}{\alpha_d+\beta_d},\\
\sigma_k^2
&=n\frac{\alpha_d\beta_d}{(\alpha_d+\beta_d)^2}
\frac{\alpha_d+\beta_d+n}{\alpha_d+\beta_d+1}.
\end{align}
The parameters $\alpha_d$ and $\beta_d$ control
the shape of the distribution, while the mean
degree depends additionally on $n$. Consequently,
varying degree heterogeneity at fixed $z$ requires
specifying how these parameters change together.
Examples at fixed mean degree are shown in the
right panel of Fig.~\ref{fig:beta-dist}.

\begin{figure}[t!]
\centering
\includegraphics[width=\linewidth]{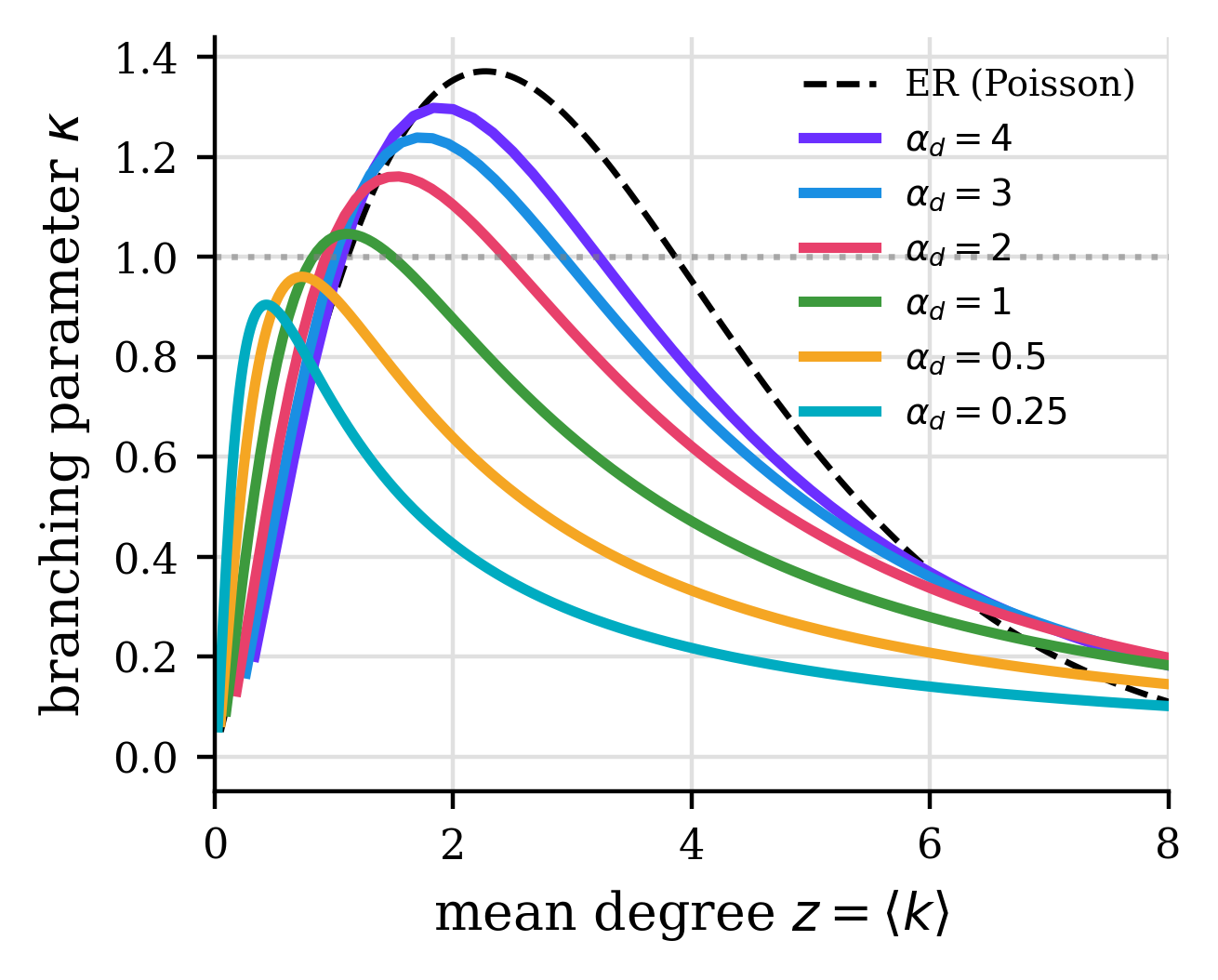}
\caption{Cascade parameter $\kappa$ as a function
of mean degree $z$ for beta-binomial degree
distributions with different values of
$\alpha_d$, fixed $\beta_d=20$, and constant
activation threshold $\bar\phi=0.25$.
Mean connectivity is varied through the maximum
degree $n$, with
$z=n\alpha_d/(\alpha_d+\beta_d)$.
The dashed black curve shows the Poisson-network
reference, while the horizontal line $\kappa=1$
marks the vulnerable-cluster onset condition.
For the parameters shown, decreasing $\alpha_d$
reduces the range of connectivity over which
$\kappa>1$.}
\label{fig:const-bb-contours}
\end{figure}

For small $\alpha_d$, suitable choices of
$\beta_d$ and $n$ produce strongly heterogeneous
degree distributions, with substantial probability
concentrated at low degrees and a broad range of
higher degrees. Increasing the concentration of
the underlying beta distribution produces more
narrowly distributed degrees. The Poisson limit
requires a joint scaling in which the beta mixing
distribution concentrates around a small
probability $z/n$, while $n\to\infty$ at fixed
$z$. Thus, the progression from heterogeneous
to Poisson-like topology depends on the chosen
parameter path rather than on $\alpha_d$ alone.

To isolate the influence of topology, we first
consider a constant activation threshold
$\bar\phi$. A node of degree $k$ can be activated
by a single neighbor only when
$1/k\geq\bar\phi$, or equivalently
$k\leq1/\bar\phi$. In the infinitesimal-seed limit, the cascade parameter therefore
reduces to
\begin{equation}
    \kappa(z)
    =
    \frac{1}{z}
    \sum_{k=2}^{\min(n,k_{\max})}
    k(k-1)P(k;n,\alpha_d,\beta_d),
    \label{eq:bb-const-kappa}
\end{equation}
where $k_{\max}=\left\lfloor\bar\phi^{-1}\right\rfloor$.

Only nodes with degrees between $2$ and
$k_{\max}$ contribute to the vulnerable
branching process. For fixed positive
$\alpha_d$ and $\beta_d$, increasing $n$
also increases the mean degree
$z=n\alpha_d/(\alpha_d+\beta_d)$.
In this large-connectivity limit, the
probability of any fixed degree satisfies
\begin{equation}
    P(k;n,\alpha_d,\beta_d)
    \sim
    n^{-\alpha_d}
    \frac{\Gamma(\alpha_d+\beta_d)}
    {\Gamma(\beta_d)}
    \frac{\Gamma(k+\alpha_d)}
    {\Gamma(\alpha_d)k!}.
\end{equation}
Consequently, for $k_{\max}\geq2$,
\begin{equation}
    \kappa(z)
    \sim
    A_d z^{-(\alpha_d+1)},
    \label{eq:bb-constant-asym}
\end{equation}
where $A_d$ is a positive constant determined
by $\alpha_d$, $\beta_d$, and $\bar\phi$.

This decay has a simple interpretation.
Under a constant fractional threshold,
nodes of sufficiently high degree cannot
be activated by a single neighbor.
As connectivity increases, the population
of low-degree vulnerable nodes becomes
increasingly rare, and their contribution
to the branching process vanishes.
Thus, sufficiently connected networks
become subcritical to infinitesimal
perturbations.

At intermediate connectivity, the
behavior depends on the degree
distribution. If $\kappa$ exceeds unity,
a finite cascade window can arise,
bounded by lower and upper connectivity
transitions. Figure~\ref{fig:const-bb-contours}
illustrates this behavior for several values
of $\alpha_d$ at fixed $\beta_d=20$.
For the parameter values shown, decreasing
$\alpha_d$ narrows the interval satisfying
$\kappa>1$ and can eliminate it entirely.
This dependence is specific to the chosen
parameterization and should not be interpreted
as a universal effect of increasing degree
heterogeneity.

\subsection{Interplay Between Degree and Threshold Heterogeneity}
\label{sec:doubly_tunable}

We now combine the beta threshold distribution of
Sec.~\ref{sec:betath} with the beta-binomial degree
distribution considered above to examine how these
two sources of heterogeneity jointly determine the
cascade condition. For independently assigned
thresholds and degrees, the cascade parameter becomes
\begin{equation}
\kappa(z)
=
\frac{1}{z}
\sum_{k=2}^{n}
k(k-1)P(k;n,\alpha_d,\beta_d)
F\left(\frac{1}{k}\right),
\label{eq:doubly_tunable}
\end{equation}
Here $z=n\alpha_d/(\alpha_d+\beta_d)$ is the
mean degree, $m=\alpha_{\rm th}/
(\alpha_{\rm th}+\beta_{\rm th})$, and
$F(x)=I_{mx/\bar\phi}(\alpha_{\rm th},\beta_{\rm th})$
within the support of the rescaled beta distribution.
The CDF saturates at unity beyond this support. Throughout, we consider the infinitesimal-seed limit.

\begin{figure*}[t]
\centering
\includegraphics[width=\textwidth]
{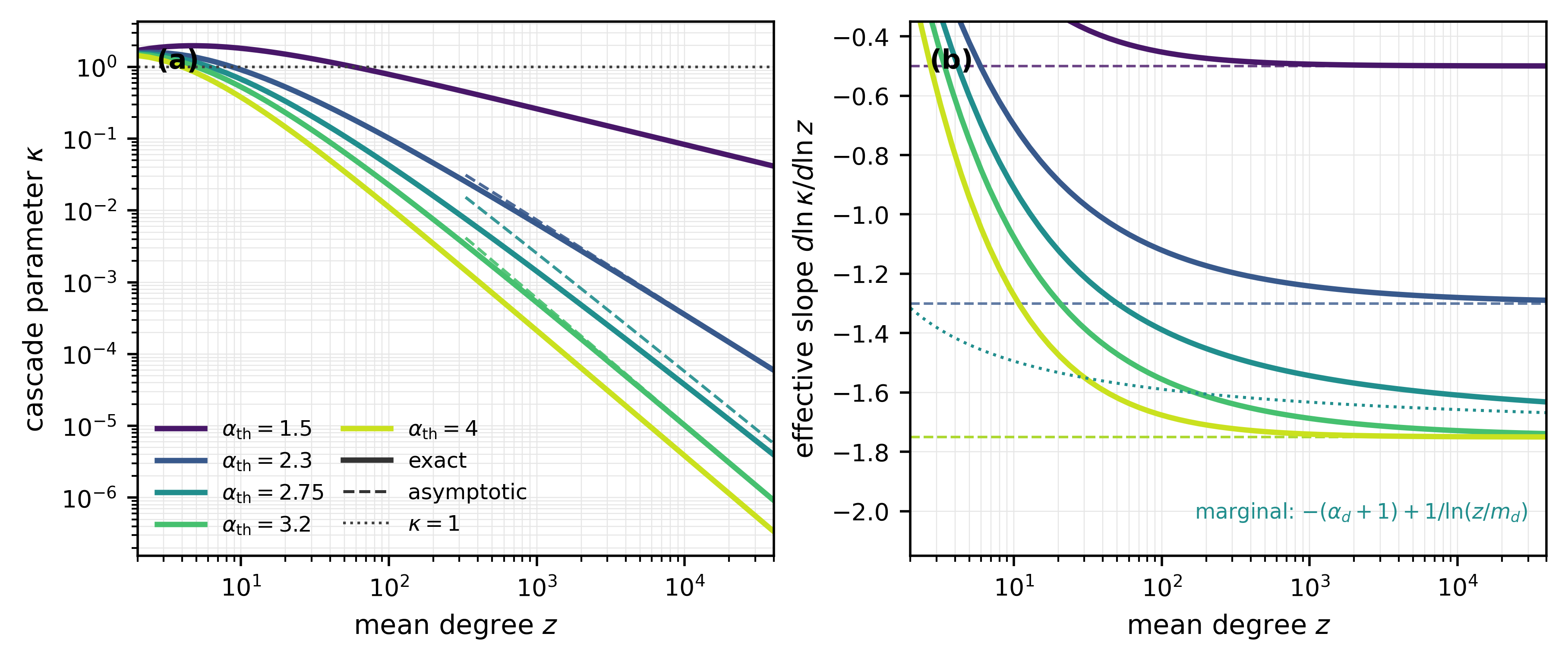}
\caption{Competition between threshold and degree
heterogeneity in the doubly tunable model.
Degrees follow the beta-binomial distribution
with $\alpha_d=0.75$ and $\beta_d=3$, while
thresholds follow the rescaled beta distribution
with $\beta_{\rm th}=3$ and mean
$\bar\phi=0.2$. The boundary between the
two asymptotic decay mechanisms is
$\alpha_{\rm th}=\alpha_d+2=2.75$.
(a) Cascade parameter $\kappa$ versus mean
degree $z$ for
$\alpha_{\rm th}=1.5,2.3,2.75,3.2,4.0$.
Solid curves are exact finite-$n$ sums,
and dashed curves are the leading
analytical asymptotes, including their
prefactors, with near perfect overlap for $\alpha_{\rm th} = 3.2$ and $4$. The dotted horizontal line
marks $\kappa=1$.
(b) Effective logarithmic slopes
$d\ln\kappa/d\ln z$ calculated from
the exact sums. Dashed horizontal lines
mark the limiting exponents
$1-\alpha_{\rm th}$ for
$\alpha_{\rm th}<\alpha_d+2$ and
$-(\alpha_d+1)$ for
$\alpha_{\rm th}\geq\alpha_d+2$.
At the marginal boundary, the dotted
curve illustrates the leading
logarithmic correction
$-(\alpha_d+1)+1/\ln(z/m_d)$.
Here $n$ is integer,
$z=nm_d$, and
$m_d=\alpha_d/(\alpha_d+\beta_d)$.}
\label{fig:bb-phase}
\end{figure*}

We examine the cascade condition as mean connectivity
increases, taking $n\to\infty$ with $\alpha_d$ and
$\beta_d$ fixed, so that $z\to\infty$. This limit
is distinct from the thermodynamic limit in the
number of vertices.

For large degrees, the threshold CDF behaves as
$F(1/k)\sim C_{\rm th}k^{-\alpha_{\rm th}}$.
The asymptotic behavior therefore depends on the
competition between the decreasing vulnerability
of highly connected nodes and the diminishing
abundance of low-degree nodes. Using the
beta-binomial mixture representation, we obtain
three regimes:
\begin{equation}
\kappa(z)\sim
\begin{cases}
C_1 z^{1-\alpha_{\rm th}},
& \alpha_{\rm th}<\alpha_d+2,\\[2mm]
C_c z^{-(\alpha_d+1)}\ln z,
& \alpha_{\rm th}=\alpha_d+2,\\[2mm]
C_2 z^{-(\alpha_d+1)},
& \alpha_{\rm th}>\alpha_d+2.
\end{cases}
\label{eq:asymptotic_regimes}
\end{equation}
The coefficients $C_1$, $C_c$, and $C_2$ are
independent of $z$. Their derivation and explicit
expressions are given in
Appendix~\ref{app:deriv}.

These regimes have distinct implications for
vulnerable branching. For $\alpha_{\rm th}<1$,
the cascade parameter grows with mean connectivity:
the increasing number of potential secondary
activations outweighs the declining vulnerability
of individual nodes. Consequently, sufficiently
connected networks satisfy $\kappa>1$ and support
a supercritical vulnerable cluster. For
$\alpha_{\rm th}>1$, the cascade parameter instead
decays, and the network eventually becomes
subcritical to infinitesimal perturbations.
At $\alpha_{\rm th}=1$, these effects balance,
and the existence of a supercritical vulnerable
cluster depends on the limiting value of $\kappa$
derived in Sec.~\ref{sec:betath}.

Within the connectivity-suppressed regime,
the boundary $\alpha_{\rm th}=\alpha_d+2$
distinguishes two asymptotic mechanisms.
For $1<\alpha_{\rm th}<\alpha_d+2$, the
dominant contribution to vulnerable branching
comes from nodes whose degrees increase with
connectivity. The threshold distribution
therefore controls the decay,
$\kappa\sim C_1z^{1-\alpha_{\rm th}}$.
For $\alpha_{\rm th}>\alpha_d+2$, nodes at
these growing degrees become too unlikely
to be vulnerable to dominate the branching
sum. The leading contribution instead comes
from relatively low-degree nodes, whose
abundance decreases as connectivity increases.
The resulting decay,
$\kappa\sim C_2z^{-(\alpha_d+1)}$, is thus
controlled by the degree distribution. At the boundary $\alpha_{\rm th}=\alpha_d+2$, vulnerable
branching is no longer dominated by either low-degree or
high-degree nodes. Instead, contributions accumulate across
a broad range of degrees, giving rise to the logarithmic
correction.

Figure~\ref{fig:bb-phase}(a) compares the exact
finite-$n$ results with the asymptotic predictions
of Eq.~\eqref{eq:asymptotic_regimes}.
The agreement at large $z$, without adjustable
parameters, confirms the predicted scaling
behavior on either side of
$\alpha_{\rm th}=\alpha_d+2$.
Figure~\ref{fig:bb-phase}(b) shows the effective
logarithmic slope $d\ln\kappa/d\ln z$, which measures
how the scaling of $\kappa$ changes with connectivity.
Below the boundary $\alpha_{\rm th}=\alpha_d+2$,
the slope approaches $1-\alpha_{\rm th}$,
while above it, the slope approaches
$-(\alpha_d+1)$, independent of the threshold
exponent. At the boundary, the logarithmic
correction leads to a much slower approach to
the limiting slope,
\begin{equation}
\frac{d\ln\kappa}{d\ln z}
\simeq
-(\alpha_d+1)+\frac{1}{\ln z}.
\label{eq:marginal_effective_slope}
\end{equation}
The dotted curve in Fig.~\ref{fig:bb-phase}(b)
shows the corresponding prediction
$-(\alpha_d+1)+1/\ln(z/m_d)$,
where $m_d=\alpha_d/(\alpha_d+\beta_d)$.

\section{Discussion and Conclusions}

Our analysis identifies two distinct boundaries governing the
interplay between degree and threshold heterogeneity in global
cascades. The first, at $\alpha_{\rm th}=1$, determines whether
increasing connectivity enhances or suppresses vulnerable
branching in the large-connectivity limit. The second, at
$\alpha_{\rm th}=\alpha_d+2$, separates two mechanisms of
suppression. Below this boundary, the asymptotic behavior is
controlled by the abundance of low-threshold individuals;
above it, the diminishing population of low-degree nodes
becomes dominant. At the boundary itself, the competition
between these contributions produces a logarithmic correction.

The physical origin of these results lies in the competing
effects of connectivity on activation and propagation.
Increasing degree provides additional paths for transmitting
activation but reduces the influence of any single neighbor.
Which effect dominates depends on how the threshold
distribution behaves near zero and how vulnerability is
distributed across degree classes. This connection between
network structure and individual response is particularly
relevant to models of social contagion, where activation often
requires reinforcement from multiple contacts. Related studies have examined how degree correlations and correlations between degree and activation thresholds influence cascade propagation~\citep{BurkholzSchweitzer2018,WuFennellPercusLerman2018}, while a
recent study of Granovetter's original model with
beta-distributed thresholds identified a corresponding
threshold exponent \cite{Fontanari2026}. Although the
microscopic dynamics differ, the appearance of the same
exponent suggests that the behavior of the threshold
distribution near zero may play a broader role in collective
activation.

The second boundary reflects the changing contribution
of different degree classes to vulnerable branching.
In the beta-binomial ensemble, increasing mean connectivity
reduces the relative abundance of low-degree nodes.
When $\alpha_{\rm th}<\alpha_d+2$, sufficiently large
degrees dominate the branching process, and its
asymptotic exponent is determined by the threshold
distribution.
When $\alpha_{\rm th}>\alpha_d+2$, vulnerability decreases
sufficiently rapidly with degree that low-degree nodes
instead dominate the sum.
Their diminishing abundance then determines the
asymptotic decay.
The logarithmic correction at
$\alpha_{\rm th}=\alpha_d+2$ marks the crossover
between these contributions.
This boundary does not separate cascading and
noncascading phases, but rather distinguishes
the mechanisms governing cascade suppression
at high connectivity.

Our choice of beta and beta-binomial distributions
provides a convenient way to examine these effects
systematically.
Their shape parameters control the abundance of
low-threshold individuals and the distribution of
connectivity, allowing us to isolate the contributions
of the two sources of heterogeneity.
While the beta-binomial has finite support for any
fixed maximum degree, our asymptotic results concern
a sequence of distributions in which this maximum
increases with mean connectivity.
Consequently, the scaling relations are properties
of the particular ensembles considered here,
rather than universal results for arbitrary degree
distributions.
Nevertheless, they demonstrate how qualitatively
different mechanisms of cascade suppression can
arise from the interaction between structural
and dynamical heterogeneity.

These results may have implications for social contagion,
where activation thresholds can represent the reinforcement
required for individuals to adopt or share information
\citep{CentolaMacy2007,Centola2010,Centola2018,
Tornberg2018,DelVicario2016,Monsted2017}.
Online platforms provide a particularly interesting
example, since recommendation algorithms can modify
patterns of exposure and interaction.
Recent work has examined how changes to content-ranking
algorithms may influence affective polarization
\citep{piccardi_reranking_2025},
while studies of echo chambers suggest that the
organization of information exchange can influence
the propagation of content
\citep{diaz_ruiz_echo_2026}.
Our results indicate that the consequences of changing
network connectivity may depend strongly on the
distribution of individual activation thresholds.
However, the present analysis does not establish
how particular interventions alter either distribution,
nor does it account for correlations between an
individual's connectivity and susceptibility to influence.
Translating these results into intervention strategies
would therefore require a more detailed model of
the underlying social dynamics.

The broader relevance of this distinction extends
beyond social contagion.
Studies of epidemic spreading have similarly demonstrated
that transmission depends on the organization of
interactions and mobility patterns, in addition to
the properties of the spreading process itself
\citep{SorianoPanos2022}. Although epidemic models generally employ different
microscopic activation rules, they illustrate the
importance of considering structural heterogeneity
alongside dynamical response when characterizing
the susceptibility of networked systems to
extensive propagation.

Our analysis is restricted to uncorrelated, locally
tree-like random networks with independently assigned
activation thresholds.
These assumptions allow us to isolate the effects
of degree and threshold heterogeneity, but exclude
several features of empirical networks.
In particular, degree correlations, clustering,
and correlations between connectivity and activation
thresholds may alter the organization of vulnerable
clusters and consequently the cascade condition \citep{DoddsPayne2009, LeeHolme2017, PayneDoddsEppstein2009}.
Furthermore, many social and biological systems
involve interactions among groups rather than
exclusively between pairs of individuals.
Random hypergraph models provide one framework
for representing such higher-order interactions
and the structural dependencies present in
empirical systems \citep{GhoshalEtAl2009}.
Recent extensions of threshold dynamics to
higher-order networks provide additional motivation
for examining how group interactions influence
cascade propagation \citep{IacopiniPetriBarratLatora2019,XuHeZhang2022,deArrudaPetriMoreno2020}.
Extending the present analysis to correlated and
higher-order networks would therefore be a natural
direction for future work.
Whether the asymptotic boundaries identified here
persist under such extensions remains an open question.

Finally, our results concern the onset of global
cascades from infinitesimal seed fractions,
as determined by the existence of a supercritical
vulnerable branching process.
They do not determine the probability that a
particular seed initiates an extensive cascade,
its eventual size, or the behavior under finite
initial perturbations.
These questions require an analysis of the full
cascade dynamics, beyond the linearized condition
considered here.
Nevertheless, the present results identify two
distinct ways in which the interaction between
network structure and individual activation thresholds
controls the propagation of small perturbations.
The distinction between threshold-controlled and
degree-controlled suppression provides a starting
point for examining these effects in more general
models of collective dynamics.

\bibliography{references}

\clearpage
\newpage 
\onecolumngrid
\appendix

\section{The beta and beta-binomial asymptotics}
\label{app:deriv}

This appendix provides the derivations of the
large-connectivity results presented in
Secs.~\ref{sec:betath}--\ref{sec:doubly_tunable}.
We consider independently assigned thresholds
and degrees in the infinitesimal-seed limit,
using the cascade parameter defined in
Eq.~\eqref{eq:kappa_intro}.

\subsection{Asymptotic behavior of the threshold distribution}

We consider a beta-distributed random variable
$X$ with shape parameters
$\alpha_{\rm th},\beta_{\rm th}>0$ and density
\begin{equation}
f_X(x)=
\frac{x^{\alpha_{\rm th}-1}
(1-x)^{\beta_{\rm th}-1}}
{B(\alpha_{\rm th},\beta_{\rm th})},
\qquad 0<x<1.
\end{equation}

To prescribe a mean threshold $\bar\phi$,
we define
$\phi=(\bar\phi/m)X$, where
$m=\alpha_{\rm th}/
(\alpha_{\rm th}+\beta_{\rm th})$.
The resulting cumulative distribution function is
\begin{equation}
F(x)=
\begin{cases}
0, & x\leq 0,\\[1mm]
I_{mx/\bar\phi}
(\alpha_{\rm th},\beta_{\rm th}),
& 0<x<\bar\phi/m,\\[1mm]
1, & x\geq\bar\phi/m.
\end{cases}
\end{equation}

Here $I_u(\alpha,\beta)$ is the regularized
incomplete beta function. As discussed in
Sec.~\ref{sec:betath}, the rescaling can produce
thresholds greater than unity, corresponding to
nodes that cannot be activated under the
fractional-threshold rule.

The behavior of the cascade parameter at large
connectivity depends on the probability that
a high-degree node is vulnerable to activation
by a single neighbor. We therefore examine
$F(1/k)$ as $k\to\infty$.

Using the integral representation
\begin{equation}
I_u(\alpha,\beta)=
\frac{1}{B(\alpha,\beta)}
\int_0^u
t^{\alpha-1}(1-t)^{\beta-1}\,dt,
\end{equation}
we expand the integrand near $t=0$ to obtain
\begin{equation}
I_u(\alpha,\beta)
=
\frac{u^\alpha}
{\alpha B(\alpha,\beta)}
\left[
1+
\frac{\alpha(1-\beta)}
{\alpha+1}u
+O(u^2)
\right].
\label{eq:Iu-asym}
\end{equation}

For a node of degree $k$, we set $x=1/k$,
so that $u=m/(k\bar\phi)$. It follows that
\begin{equation}
F\left(\frac{1}{k}\right)
=
C_{\rm th}k^{-\alpha_{\rm th}}
\left[1+O(k^{-1})\right],
\label{eq:Ftail}
\end{equation}
where
\begin{equation}
C_{\rm th}
=
\frac{m^{\alpha_{\rm th}}}
{\alpha_{\rm th}
B(\alpha_{\rm th},\beta_{\rm th})
\bar\phi^{\alpha_{\rm th}}}.
\end{equation}

Thus, the exponent $\alpha_{\rm th}$
determines how single-neighbor vulnerability
decays with degree, while the remaining
threshold parameters enter through the
prefactor $C_{\rm th}$.

\subsection{The cascade parameter on
Erd\H{o}s--R\'enyi graphs}

For an Erd\H{o}s--R\'enyi network,
the degree distribution is Poisson,
\begin{equation}
p_k=e^{-z}\frac{z^k}{k!},
\end{equation}
where $z$ is the mean degree.

Substituting into Eq.~\eqref{eq:kappa_intro},
we obtain
\begin{equation}
\begin{aligned}
\kappa(z)
&=
\frac{1}{z}
\sum_{k=2}^{\infty}
k(k-1)e^{-z}\frac{z^k}{k!}
F\left(\frac{1}{k}\right)\\
&=
e^{-z}
\sum_{k=2}^{\infty}
\frac{z^{k-1}}{(k-2)!}
F\left(\frac{1}{k}\right).
\end{aligned}
\end{equation}

At large mean degree, the Poisson
distribution becomes concentrated around
$k\sim z$. Using Eq.~\eqref{eq:Ftail},
we therefore obtain the leading behavior
\begin{equation}
\kappa(z)
\sim
C_{\rm th}z^{1-\alpha_{\rm th}}.
\label{eq:er-asym-appendix}
\end{equation}

The sign of $1-\alpha_{\rm th}$ determines
the large-connectivity behavior.
For $\alpha_{\rm th}<1$, the cascade
parameter grows with $z$, whereas for
$\alpha_{\rm th}>1$, it decreases.

At $\alpha_{\rm th}=1$, the leading
power vanishes, and the limiting value
must be determined from the prefactor.
In this case,
\begin{equation}
I_u(1,\beta_{\rm th})
=
1-(1-u)^{\beta_{\rm th}}
=
\beta_{\rm th}u+O(u^2).
\end{equation}

Since
$m=1/(1+\beta_{\rm th})$, we have
\begin{equation}
F\left(\frac{1}{k}\right)
=
\frac{\beta_{\rm th}}
{(1+\beta_{\rm th})\bar\phi}
\frac{1}{k}
+O(k^{-2}).
\end{equation}

Substitution into the Poisson cascade
parameter gives
\begin{equation}
\kappa(z)
=
\frac{\beta_{\rm th}}
{(1+\beta_{\rm th})\bar\phi}
\left[
1+O(z^{-1})
\right].
\end{equation}

Consequently,
\begin{equation}
\kappa_\infty
=
\lim_{z\to\infty}\kappa(z)
=
\frac{\beta_{\rm th}}
{(1+\beta_{\rm th})\bar\phi}.
\label{eq:kappainf-univ}
\end{equation}

The critical mean threshold at this
boundary follows from
$\kappa_\infty=1$:
\begin{equation}
\bar\phi_{\rm crit}
=
\frac{\beta_{\rm th}}
{1+\beta_{\rm th}}.
\end{equation}

This establishes the
$\alpha_{\rm th}=1$ boundary for
Erd\H{o}s--R\'enyi networks.
The corresponding behavior for the
beta-binomial topology follows from
the asymptotic analysis below.

\subsection{The beta-binomial degree distribution
in the large-connectivity limit}

The beta-binomial distribution admits a useful
representation as a mixture of binomial distributions.
Let $p$ be drawn from a beta distribution,
\begin{equation}
f_d(p)=
\frac{p^{\alpha_d-1}(1-p)^{\beta_d-1}}
{B(\alpha_d,\beta_d)},
\qquad 0<p<1,
\label{eq:beta_mixing}
\end{equation}
and, conditional on $p$, let the degree $K$ be
distributed as $\mathrm{Binomial}(n,p)$.
Integrating over $p$ gives
\begin{equation}
\begin{aligned}
P(k;n,\alpha_d,\beta_d)
&=
\int_0^1
\binom{n}{k}p^k(1-p)^{n-k}f_d(p)\,dp\\
&=
\binom{n}{k}
\frac{B(k+\alpha_d,n-k+\beta_d)}
{B(\alpha_d,\beta_d)}.
\end{aligned}
\label{eq:bb_mixture}
\end{equation}

The mean degree is
\begin{equation}
z=\langle K\rangle
=n m_d,
\qquad
m_d=\frac{\alpha_d}{\alpha_d+\beta_d}.
\label{eq:bb_mean_appendix}
\end{equation}

We study the large-connectivity limit by taking
$n\to\infty$ at fixed positive $\alpha_d$ and
$\beta_d$, so that $z\to\infty$.
This should be distinguished from the thermodynamic
limit $N\to\infty$, where $N$ denotes the number
of vertices in the network.

For fixed $p>0$, the conditional degree
distribution becomes concentrated around $K=np$.
Consequently,
\begin{equation}
\frac{K}{n}\longrightarrow p
\label{eq:conditional_concentration}
\end{equation}
in probability as $n\to\infty$.

This representation allows us to calculate the
large-connectivity cascade parameter directly,
without approximating the beta-binomial distribution
by a power law with an exponential cutoff.

\subsection{Asymptotic behavior of the cascade parameter}

In the infinitesimal-seed limit, the cascade
parameter is
\begin{equation}
\kappa(z)=
\frac{1}{z}
\sum_{k=2}^{n}
k(k-1)P(k;n,\alpha_d,\beta_d)
F\left(\frac{1}{k}\right).
\label{eq:kappa_bb_appendix}
\end{equation}

Using the large-degree behavior of the rescaled-beta
threshold distribution derived in Eq.~\eqref{eq:Ftail},
we have
\begin{equation}
F\left(\frac{1}{k}\right)
\sim C_{\rm th}k^{-\alpha_{\rm th}},
\qquad k\to\infty,
\label{eq:threshold_tail_appendix}
\end{equation}
where
\begin{equation}
C_{\rm th}=
\frac{m^{\alpha_{\rm th}}}
{\alpha_{\rm th}
B(\alpha_{\rm th},\beta_{\rm th})
\bar{\phi}^{\alpha_{\rm th}}},
\qquad
m=\frac{\alpha_{\rm th}}
{\alpha_{\rm th}+\beta_{\rm th}}.
\label{eq:Cth_appendix}
\end{equation}

We distinguish three regimes according to the
relative values of $\alpha_{\rm th}$ and
$\alpha_d+2$.

\subsubsection{Threshold-dominated regime:
$\alpha_{\rm th}<\alpha_d+2$}

For large degrees,
\begin{equation}
k(k-1)F(1/k)
\sim C_{\rm th}k^{2-\alpha_{\rm th}}.
\end{equation}

Using the mixture representation, the leading
behavior of the cascade parameter is therefore
\begin{equation}
\kappa(z)
\sim
\frac{C_{\rm th}}{z}
n^{2-\alpha_{\rm th}}
\int_0^1 p^{2-\alpha_{\rm th}}
f_d(p)\,dp.
\label{eq:kappa_mixture_moment}
\end{equation}

The integral is a moment of the beta distribution,
\begin{equation}
\begin{aligned}
\int_0^1 p^{2-\alpha_{\rm th}}f_d(p)\,dp
&=
\frac{1}{B(\alpha_d,\beta_d)}
\int_0^1
p^{\alpha_d+1-\alpha_{\rm th}}
(1-p)^{\beta_d-1}\,dp\\
&=
\frac{
B(\alpha_d+2-\alpha_{\rm th},\beta_d)}
{B(\alpha_d,\beta_d)}.
\end{aligned}
\label{eq:beta_moment}
\end{equation}

This moment is finite when
\begin{equation}
\alpha_{\rm th}<\alpha_d+2.
\label{eq:moment_condition}
\end{equation}

In this regime, the leading contribution comes
from degrees that grow proportionally with $n$.
The contribution from degrees of order unity
is asymptotically smaller, so the large-degree
threshold expansion can be used in the leading
term.

Substituting $n=z/m_d$ into
Eq.~\eqref{eq:kappa_mixture_moment} gives
\begin{equation}
\kappa(z)\sim
C_1z^{1-\alpha_{\rm th}},
\label{eq:threshold_dominated_appendix}
\end{equation}
where
\begin{equation}
C_1=
C_{\rm th}m_d^{\alpha_{\rm th}-2}
\frac{
B(\alpha_d+2-\alpha_{\rm th},\beta_d)}
{B(\alpha_d,\beta_d)}.
\label{eq:C1_appendix}
\end{equation}

Thus the threshold exponent determines the
large-connectivity scaling, while the topology
enters through a multiplicative constant.

\subsubsection{Degree-dominated regime:
$\alpha_{\rm th}>\alpha_d+2$}

When $\alpha_{\rm th}>\alpha_d+2$, the beta
moment in Eq.~\eqref{eq:beta_moment} diverges.
The leading contribution is instead controlled
by degrees that remain finite as $n$ increases.

For fixed $k$, the exact beta-binomial
distribution satisfies
\begin{equation}
\begin{aligned}
P(k;n,\alpha_d,\beta_d)
&=
\frac{\Gamma(n+1)}
{\Gamma(k+1)\Gamma(n-k+1)}
\\
&\quad\times
\frac{
\Gamma(k+\alpha_d)
\Gamma(n-k+\beta_d)
\Gamma(\alpha_d+\beta_d)}
{
\Gamma(n+\alpha_d+\beta_d)
\Gamma(\alpha_d)\Gamma(\beta_d)
}.
\end{aligned}
\label{eq:bb_gamma_form}
\end{equation}

Using the large-$n$ asymptotic ratios of gamma
functions, we obtain
\begin{equation}
P(k;n,\alpha_d,\beta_d)
\sim
n^{-\alpha_d}
\frac{\Gamma(\alpha_d+\beta_d)}
{\Gamma(\beta_d)}
\frac{\Gamma(k+\alpha_d)}
{\Gamma(\alpha_d)k!}.
\label{eq:bb_fixed_degree}
\end{equation}

Substituting into the cascade parameter gives
\begin{equation}
\kappa(z)\sim
C_2z^{-(\alpha_d+1)},
\label{eq:degree_dominated_appendix}
\end{equation}
with
\begin{equation}
C_2=
\frac{\Gamma(\alpha_d+\beta_d)}
{\Gamma(\beta_d)}
m_d^{\alpha_d}
\sum_{k=2}^{\infty}
\frac{\Gamma(k+\alpha_d)}
{\Gamma(\alpha_d)k!}
k(k-1)F\left(\frac{1}{k}\right).
\label{eq:C2_appendix}
\end{equation}

The sum converges because, for large $k$,
its summand behaves as
\begin{equation}
k^{1+\alpha_d-\alpha_{\rm th}},
\end{equation}
which is summable when
$\alpha_{\rm th}>\alpha_d+2$.

Importantly, the coefficient $C_2$ depends
on the exact threshold probabilities
$F(1/k)$ at finite degrees, rather than
only on the asymptotic coefficient
$C_{\rm th}$. The exponent, however, is
determined entirely by the degree parameter
$\alpha_d$.

The contribution from degrees proportional
to $n$ scales as $z^{1-\alpha_{\rm th}}$,
which is subleading in this regime.

\subsubsection{Marginal regime:
$\alpha_{\rm th}=\alpha_d+2$}

At the boundary between the two regimes,
the asymptotic summand in
Eq.~\eqref{eq:C2_appendix} behaves as $1/k$.
The corresponding partial sum diverges
logarithmically,
\begin{equation}
\sum_{k\lesssim n}\frac{1}{k}
\sim\ln n.
\end{equation}

The cascade parameter therefore has the
marginal scaling
\begin{equation}
\kappa(z)\sim
C_cz^{-(\alpha_d+1)}\ln z,
\label{eq:marginal_appendix}
\end{equation}
where
\begin{equation}
C_c=
C_{\rm th}
\frac{m_d^{\alpha_d}}
{B(\alpha_d,\beta_d)}.
\label{eq:Cc_appendix}
\end{equation}

The logarithmic correction reflects the
marginal convergence of the degree-weighted
vulnerability sum at
$\alpha_{\rm th}=\alpha_d+2$.

Together, Eqs.~\eqref{eq:threshold_dominated_appendix},
\eqref{eq:degree_dominated_appendix}, and
\eqref{eq:marginal_appendix} establish the
three asymptotic regimes stated in
Eq.~\eqref{eq:asymptotic_regimes}.

\end{document}